\documentclass[aps,prl,twocolumn,superscriptaddress]{revtex4}

\usepackage[dvips]{graphicx}

\begin{document}

\title{Extending the search for neutrino point sources with IceCube above the horizon}

\affiliation{III. Physikalisches Institut, RWTH Aachen University, D-52056 Aachen, Germany}
\affiliation{Dept.~of Physics and Astronomy, University of Alabama, Tuscaloosa, AL 35487, USA}
\affiliation{Dept.~of Physics and Astronomy, University of Alaska Anchorage, 3211 Providence Dr., Anchorage, AK 99508, USA}
\affiliation{CTSPS, Clark-Atlanta University, Atlanta, GA 30314, USA}
\affiliation{School of Physics and Center for Relativistic Astrophysics, Georgia Institute of Technology, Atlanta, GA 30332. USA}
\affiliation{Dept.~of Physics, Southern University, Baton Rouge, LA 70813, USA}
\affiliation{Dept.~of Physics, University of California, Berkeley, CA 94720, USA}
\affiliation{Lawrence Berkeley National Laboratory, Berkeley, CA 94720, USA}
\affiliation{Institut f\"ur Physik, Humboldt-Universit\"at zu Berlin, D-12489 Berlin, Germany}
\affiliation{Fakult\"at f\"ur Physik \& Astronomie, Ruhr-Universit\"at Bochum, D-44780 Bochum, Germany}
\affiliation{Physikalisches Institut, Universit\"at Bonn, Nussallee 12, D-53115 Bonn, Germany}
\affiliation{Universit\'e Libre de Bruxelles, Science Faculty CP230, B-1050 Brussels, Belgium}
\affiliation{Vrije Universiteit Brussel, Dienst ELEM, B-1050 Brussels, Belgium}
\affiliation{Dept.~of Physics, Chiba University, Chiba 263-8522, Japan}
\affiliation{Dept.~of Physics and Astronomy, University of Canterbury, Private Bag 4800, Christchurch, New Zealand}
\affiliation{Dept.~of Physics, University of Maryland, College Park, MD 20742, USA}
\affiliation{Dept.~of Physics and Center for Cosmology and Astro-Particle Physics, Ohio State University, Columbus, OH 43210, USA}
\affiliation{Dept.~of Astronomy, Ohio State University, Columbus, OH 43210, USA}
\affiliation{Dept.~of Physics, TU Dortmund University, D-44221 Dortmund, Germany}
\affiliation{Dept.~of Subatomic and Radiation Physics, University of Gent, B-9000 Gent, Belgium}
\affiliation{Max-Planck-Institut f\"ur Kernphysik, D-69177 Heidelberg, Germany}
\affiliation{Dept.~of Physics and Astronomy, University of California, Irvine, CA 92697, USA}
\affiliation{Laboratory for High Energy Physics, \'Ecole Polytechnique F\'ed\'erale, CH-1015 Lausanne, Switzerland}
\affiliation{Dept.~of Physics and Astronomy, University of Kansas, Lawrence, KS 66045, USA}
\affiliation{Dept.~of Astronomy, University of Wisconsin, Madison, WI 53706, USA}
\affiliation{Dept.~of Physics, University of Wisconsin, Madison, WI 53706, USA}
\affiliation{Institute of Physics, University of Mainz, Staudinger Weg 7, D-55099 Mainz, Germany}
\affiliation{University of Mons-Hainaut, 7000 Mons, Belgium}
\affiliation{Bartol Research Institute and Department of Physics and Astronomy, University of Delaware, Newark, DE 19716, USA}
\affiliation{Dept.~of Physics, University of Oxford, 1 Keble Road, Oxford OX1 3NP, UK}
\affiliation{Dept.~of Physics, University of Wisconsin, River Falls, WI 54022, USA}
\affiliation{Oskar Klein Centre and Dept.~of Physics, Stockholm University, SE-10691 Stockholm, Sweden}
\affiliation{Dept.~of Astronomy and Astrophysics, Pennsylvania State University, University Park, PA 16802, USA}
\affiliation{Dept.~of Physics, Pennsylvania State University, University Park, PA 16802, USA}
\affiliation{Dept.~of Physics and Astronomy, Uppsala University, Box 516, S-75120 Uppsala, Sweden}
\affiliation{Dept.~of Physics and Astronomy, Utrecht University/SRON, NL-3584 CC Utrecht, The Netherlands}
\affiliation{Dept.~of Physics, University of Wuppertal, D-42119 Wuppertal, Germany}
\affiliation{DESY, D-15735 Zeuthen, Germany}

\author{R.~Abbasi}
\affiliation{Dept.~of Physics, University of Wisconsin, Madison, WI 53706, USA}
\author{Y.~Abdou}
\affiliation{Dept.~of Subatomic and Radiation Physics, University of Gent, B-9000 Gent, Belgium}
\author{T.~Abu-Zayyad}
\affiliation{Dept.~of Physics, University of Wisconsin, River Falls, WI 54022, USA}
\author{J.~Adams}
\affiliation{Dept.~of Physics and Astronomy, University of Canterbury, Private Bag 4800, Christchurch, New Zealand}
\author{J.~A.~Aguilar}
\affiliation{Dept.~of Physics, University of Wisconsin, Madison, WI 53706, USA}
\author{M.~Ahlers}
\affiliation{Dept.~of Physics, University of Oxford, 1 Keble Road, Oxford OX1 3NP, UK}
\author{K.~Andeen}
\affiliation{Dept.~of Physics, University of Wisconsin, Madison, WI 53706, USA}
\author{J.~Auffenberg}
\affiliation{Dept.~of Physics, University of Wuppertal, D-42119 Wuppertal, Germany}
\author{X.~Bai}
\affiliation{Bartol Research Institute and Department of Physics and Astronomy, University of Delaware, Newark, DE 19716, USA}
\author{M.~Baker}
\affiliation{Dept.~of Physics, University of Wisconsin, Madison, WI 53706, USA}
% ...

\date{\today}

\collaboration{IceCube Collaboration}
\noaffiliation

\begin{abstract}
Point source searches with the IceCube neutrino telescope have been restricted to one hemisphere, due to the exclusive selection of upward going events as a way of rejecting the atmospheric muon background. 
We show that the region above the horizon can be included by suppressing the background through energy-sensitive cuts. 
This approach improves the sensitivity above PeV energies, previously not accessible for declinations of more than a few degrees below the horizon due to the absorption of neutrinos in Earth.
We present results based on data collected with 22 strings of IceCube, extending its field of view and energy reach for point source searches. No significant excess above the atmospheric background is observed 
in a sky scan and in tests of source candidates. Upper limits are reported, which for the first time cover point sources in the southern sky up to EeV energies.
\end{abstract}


\maketitle

\begin{figure}[t]
\includegraphics[width=246pt]{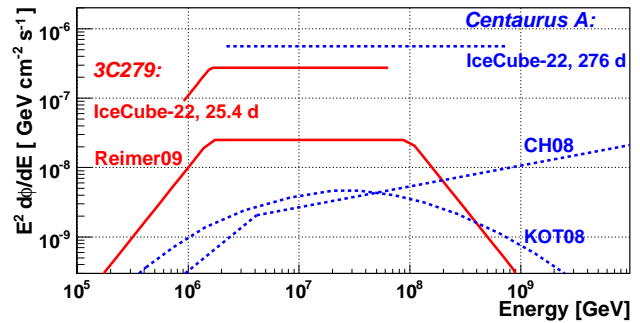}%
\caption{\label{fig_lim}(Color online). Flux limits from this work for 2 sources: Centaurus A limit (blue dashed) with models CH08~\cite{Cuoco} and KOT08~\cite{Kachelriess}; 3C279 flare search (red solid) with model Reimer09~\cite{reimer2009hd}.}
\end{figure}

The spectrum of charged cosmic rays is known to extend to energies larger than $10^{20}$ eV, while the sources and acceleration processes leading to their production remain unclear. 
Active Galactic Nuclei (AGN) are possible source candidates and results from the Pierre Auger Observatory~\cite{Auger} highlight, in particular, Centaurus A as an AGN near a cosmic ray hot spot in the southern hemisphere. Neutrinos up to EeV energies could be produced in hadronic acceleration processes at this site, predicted for example in~\cite{Cuoco}. With the possibility to backtrace these unique messengers, which propagate through space practically without deflection or absorption, neutrino telescopes like IceCube \cite{icecube} have the opportunity to identify such sources.
\\
The search for neutrino point sources relies on observing and reconstructing muon track directions inside the detector volume. 
Expected signals stem from charged-current interactions of muon neutrinos. The contribution from tau neutrinos is small~\cite{Beacom2002} and not included in the calculations presented here. 
In IceCube, Cherenkov light from these muon tracks is detected in digital optical modules (DOM)~\cite{dom}, with 60 of these arranged vertically on a string and a total of 80 strings to be installed by 2011 in the deep ice at the geographic South Pole.
\\
The background consists primarily of high energy muons, produced in extended air showers, and a much lower contribution of atmospheric neutrinos from the same interactions.
A standard method to suppress the muons from above the horizon is a cut on the reconstructed track direction as in \cite{ic22ps09}, which excludes the southern hemisphere from IceCube neutrino point source searches. 
It also limits the observable part of the spectrum, since the neutrino-nucleus cross section rises with energy and thus leads to absorption of neutrinos going up through Earth with energies larger than $\sim 1$~PeV.
To extend the reach of a neutrino analysis to higher energies and enlarge the field of view, tracks from above the horizon have to be included.
Presented here is the first point source analysis which achieves this through suppression of lower energy background muons. This sets it apart from muon astronomy like in \cite{Ahlen1993} and makes this search sensitive to cosmic neutrinos which are expected to follow a harder spectrum than the atmospheric background.
A possible signal contribution from gamma-ray showers from above the horizon cannot be excluded~\cite{Halzen2003}, but at PeV energies this could only be expected from very close sources.
The event selection described in the following was optimized for a binned study of track directions and requires harder cuts than the likelihood analysis in~\cite{ic22ps09} to achieve the best sensitivity.
\\
Data collected with 22 strings of IceCube were analyzed for this work, summing to a total of 276 days livetime over 310 days between May 2007 and April 2008. 
An online filter provided an all-sky event sample with a rate of $23.6$ Hz.
An elaborate direction reconstruction based on likelihood maximization was performed offline.
The extended probability density function used in this work describes the detection of first photon pulses in a DOM, taking into account the total number of photoelectrons instead of ignoring all later hits as was done in previous IceCube searches.
Both methods are described in~\cite{Ahrens2004, Japaridze2005}. 
The enhanced reconstruction approach leads to an improved angular resolution at higher energies, due to the importance of additional Cherenkov light from the increasing number of secondary particles along the track.
The median angular resolution for the final event selection described below is $1.3^{\circ}$ ($1.2^{\circ}$) under the assumption of an $E^{-2}$ ($E^{-1.5}$) spectrum.
\begin{figure}
\includegraphics[width=246pt]{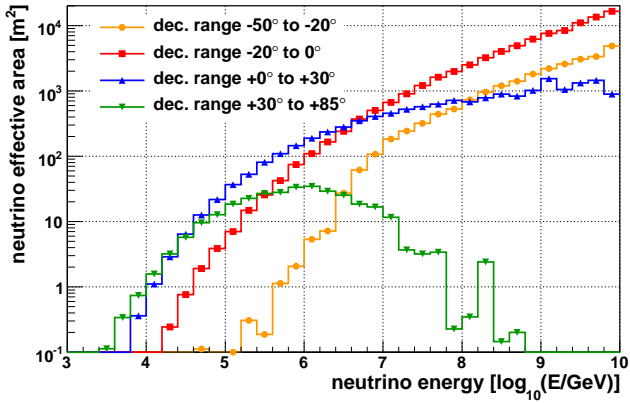}%
\caption{\label{fig_aeff}(Color online). $\nu_{\mu}+\bar{\nu}_{\mu}$ effective areas
for this analysis with 22 strings. 
At positive declinations, highest energies are suppressed through absorption. 
The line for $-50^{\circ}$ to $-20^{\circ}$ is lowered due to strong cuts above the horizon.
}
\end{figure}
\begin{figure}
\includegraphics[width=246pt]{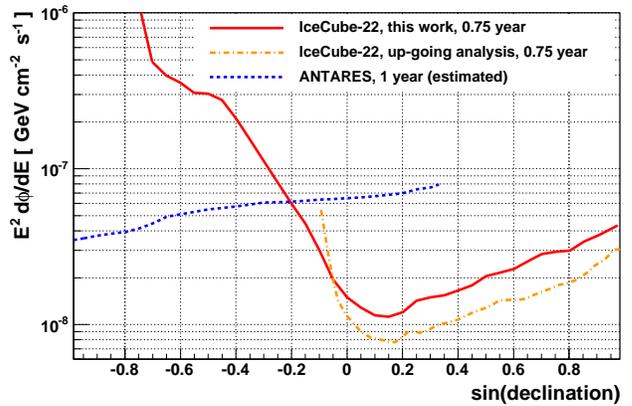}
\caption{\label{fig_sens}(Color online). Sensitivity to point sources with $E^{-2}$ $\nu_{\mu}$+$\bar{\nu}_{\mu}$ spectra vs. declination, comparing this analysis with \cite{ic22ps09} and ANTARES~\cite{Antares}.
Differences in the energy coverage at southern declinations are discussed in the text and Fig.~\ref{fig_map}.
}
\end{figure}
\\
Only events with declinations $\geq\!\!-50^{\circ}$ were considered in the analysis, due to the very low signal expectations from the small interaction volume directly above the detector.
The atmospheric background rises steeply with the angle above the horizon and is dominated by bundles of multiple muons. 
Their combined Cherenkov emission has a signature similar to a single very high energy muon track induced by a neutrino.
A cut based on the number of photon pulses per DOM, scaled with the reconstructed declination and estimated energy, was applied to suppress this background topology.
\\
Further background suppression was necessary and followed the approach from~\cite{Achterberg:2006vc} with declination dependent event selection criteria.
Cuts on five parameters were simultaneously optimized for the best sensitivity to two template signal spectra, a standard $E^{-2}$ and a harder $E^{-1.5}$. 
The cut variables included two energy estimators which were crucial for background rejection through rising energy thresholds above the horizon.
For upward going events, the efficient signal separation relied mainly on track quality parameters to reject misreconstructed muons. 
Final cuts were parametrized as linear functions of declination, smoothly connecting both hemispheres.
\\
The resulting sample consists of 1877 events with declinations in the range $-50^{\circ}$ to $+85^{\circ}$. Simulations show that tracks coming from below are mainly induced by atmospheric neutrinos, while downward going events are dominated by atmospheric muons.
The comparison of effective areas in Fig.~\ref{fig_aeff} illustrates the fact that neutrinos at highest energies are absorbed prior to detection when coming from too far below the horizon, but are accessible as downward going tracks, as discussed above.
\\
\begin{figure*}
\includegraphics[width=492pt]{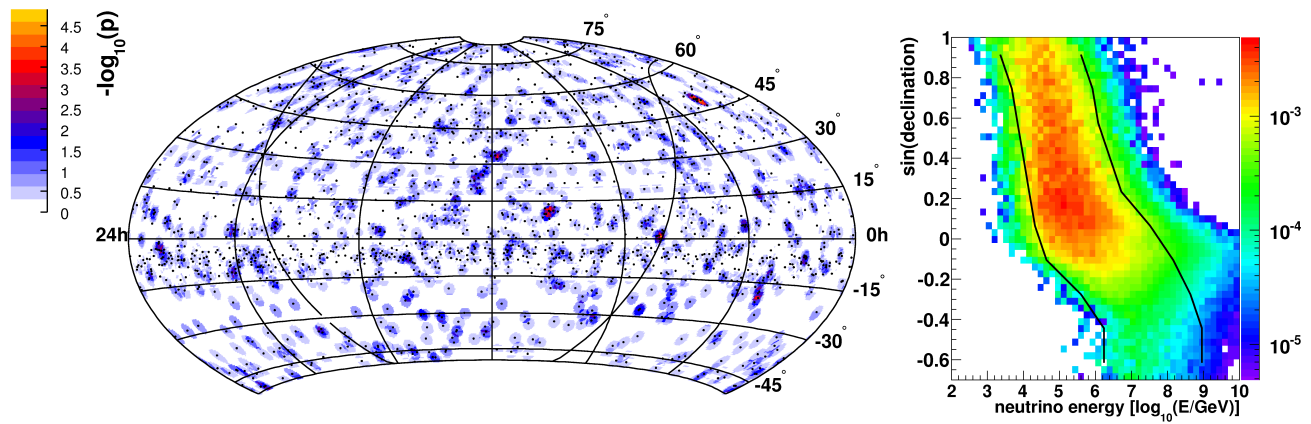} 
\caption{\label{fig_map}(Color online). Left: Sky map of the directions of the 1877 events in equatorial coordinates. The color indicates the negative logarithm of the p-value for compatibility with background in each angular search bin before trial-corrections. Right: Energies for a simulated $E^{-2}$ neutrino spectrum as a function of declination. The color indicates the fraction of events per angle and energy bin at final cut level. The black lines describe the region of 90\% signal containment in each declination band.}
\end{figure*}
The reconstructed event directions were kept blinded for the optimization of the analysis in order to avoid bias. In a final step, the true astronomical coordinates were reintroduced to determine significances for the scan of the sky and tests of candidate sources, described in the following.
Details on the binned search for spatial clustering are given in~\cite{Achterberg:2006vc}. The major difference is that in this work we extended the scan to the southern sky.
A declination dependent optimization of the search bin radius showed only a weak dependence of the sensitivity on the exact size and led to a fixed value of $2.5^{\circ}$.
The sensitivity to an $E^{-2}$ spectrum, defined as the mean upper limit at 90\% confidence level constructed as in~\cite{feldman1998}, is displayed in Fig.~\ref{fig_sens}.
The steep rise in the southern hemisphere is due to the increasing energy-sensitive cuts above the horizon. Signal efficiency depends strongly on declination and effectively defines a lower energy threshold rising from the TeV regime in the North to PeV energies in the South for an $E^{-2}$ spectrum.
The sensitivity can be compared to projections for ANTARES~\cite{Antares}, which mainly covers GeV to TeV energies when looking down to the southern hemisphere.
Since the energy range for IceCube is shifted well into the PeV regime above the horizon, both experiments are complementary in covering the full neutrino spectrum.
\\
No significant deviation from the background hypothesis was found in the sky map, shown in Fig.~\ref{fig_map}. 
The fluctuation with the highest significance is in the direction of $103.5^{\circ}$ r.a.~and $1.0^{\circ}$ dec.~(8 events observed, 1.2 expected).
By performing searches on randomized sky maps, we calculated the (post-trial) probability for an equally or more significant excess anywhere in the sky due to a background fluctuation to be 37\%.
\begin{table}[t]
\caption{ \label{sourcelist}(color online). Results for pre-defined source candidates in the southern hemisphere. $\mu_{90}$ is the neutrino event upper limit of the Feldman-Cousins 90\% confidence interval, $\Phi_{90}$ the resulting flux upper limit for an $E^{-2}$ neutrino spectrum, i.e. $d\Phi/dE\leq \Phi_{90} 10^{-9} \text{GeV}^{-1}\text{cm}^{-2}\text{s}^{-1}(E/\text{GeV})^{-2}$. $\Delta E$ is the energy range of 90\% signal containment.}
\begin{ruledtabular}
\begin{tabular}{@{}lccccc@{}} \toprule

Object & r.a. [$^{\circ}$] & dec. [$^{\circ}$] & $\mu_{90}$ & $\Phi_{90}$ & $\Delta E$ [GeV] \\ \colrule
 PKS 0537-441 & 84.7 & $-44.1$ & 1.8 & 281.3 & $3\!\!\times\!\!10^{6}$ -- $7\!\!\times\!\!10^{8}$ \\
 Centaurus A & 201.4 & $-43.0$ & 3.7 & 556.7 & $2\!\!\times\!\!10^{6}$ -- $8\!\!\times\!\!10^{8}$ \\
 PKS 1454-354 & 224.4 & $-35.6$ & 2.0 & 238.9 & $1\!\!\times\!\!10^{6}$ -- $9\!\!\times\!\!10^{8}$ \\
 PKS 1622-297 & 246.5 & $-29.9$ & 4.0 & 442.8 & $1\!\!\times\!\!10^{6}$ -- $9\!\!\times\!\!10^{8}$ \\
 Sgr A$^*$ & 266.4 & $-29.0$ & 2.3 & 244.3 & $1\!\!\times\!\!10^{6}$ -- $9\!\!\times\!\!10^{8}$ \\ 
 PKS 1622-253 & 246.4 & $-25.5$ & 2.4 & 218.3 & $1\!\!\times\!\!10^{6}$ -- $8\!\!\times\!\!10^{8}$ \\
 PKS 1830-21 & 278.4 & $-21.1$ & 2.3 & 135.5 & $8\!\!\times\!\!10^{5}$ -- $6\!\!\times\!\!10^{8}$ \\
 PKS 1730-130  & 263.3 & $-13.1$ & 4.6 & 92.4 & $2\!\!\times\!\!10^{5}$ -- $3\!\!\times\!\!10^{8}$ \\
 PKS 1510-089 & 228.2 & $-9.1$ & 3.4 & 38.3 & $8\!\!\times\!\!10^{4}$ -- $2\!\!\times\!\!10^{8}$ \\
 PKS 1406-076 & 212.2 & $-7.9$ & 4.2 & 38.2 & $6\!\!\times\!\!10^{4}$ -- $2\!\!\times\!\!10^{8}$ \\
 PKS 2023-07 & 306.4 & $-7.6$ & 2.8 & 24.8 & $6\!\!\times\!\!10^{4}$ -- $2\!\!\times\!\!10^{8}$ \\
 3C279  & 194.1 & $-5.8$ & 4.0 & 27.1 & $5\!\!\times\!\!10^{4}$ -- $1\!\!\times\!\!10^{8}$ \\
 PKS 0336-019  & 54.9 & $-1.8$ & 3.0 & 14.7 & $3\!\!\times\!\!10^{4}$ -- $9\!\!\times\!\!10^{7}$ \\
 PKS 0420-014 & 65.8 & $-1.3$ & 3.1 & 14.7 & $2\!\!\times\!\!10^{4}$ -- $8\!\!\times\!\!10^{7}$ \\

\botrule
 \end{tabular}
 \end{ruledtabular}
\end{table}
\\
We performed a separate test of pre-defined source candidates, aiming at reducing the effects of trial factors.
The selection is focused on AGN with high fluxes of GeV gamma-rays, motivated by hadronic models like \cite{Atoyan2004}. 
By imposing thresholds for average (maximum) photon fluxes of $15\!\times\!10^{-8}$ cm$^{-2}$s$^{-1}$ ($40\!\times\!10^{-8}$ cm$^{-2}$s$^{-1}$), we selected 23 bright blazars from the confirmed AGN in the third EGRET catalog \cite{hartman1999}, the most complete list of GeV detections at the time our list was defined.
Two additional blazars reported by Fermi in November 2008 \cite{2008ATel1701,2008ATel1650} were also included based on the same criteria, as well as the close-by AGN M87. 
With the extension to the southern hemisphere, this is the first IceCube analysis that allows a search for neutrinos from Centaurus A and the center of our galaxy, Sgr A$^*$, and both were added to the list.
\\
No significant excess above background was found for any of the candidates. The upward fluctuation with the highest significance occurs for PKS 1622-297 with 1 event observed and 0.3 expected. From background simulations, we expect a similar or more significant result from any of the sources in 98\% of all cases. Flux limits, calculated according to \cite{feldman1998}, are shown in Table~\ref{sourcelist} for candidates in the southern sky. 
They are considered to be valid in the given energy ranges containing 90\% of the signal expectation.
For sources in the northern hemisphere, the dedicated IceCube analysis in~\cite{ic22ps09} provides better limits.
The flux limit obtained for Centaurus A is shown in Fig.~\ref{fig_lim} in relation to neutrino predictions from Cuoco \& Hannestad \cite{Cuoco} and Kachelriess et al. \cite{Kachelriess}.
\\
For the blazar 3C279, we considered the possibility of variable
emission, motivated by the reports on a gamma-ray flare by MAGIC~\cite{magic3C279}. Since hadronic acceleration models seem to be favored for this event~\cite{boettcher2008}, we performed a time dependent test of neutrino fluxes predicted in \cite{reimer2009hd}.
The most promising periods for flares during the livetime for the analysis were selected based on X-ray and optical observations from \cite{Larionov2008}, requiring enhanced emission in both wavebands. The chosen periods, given in MJDs, are 54258.6 to 54260.6, 54307 to 54314.5, 54324.1 to 54339.9~and 54348 to 54350, adding up to a total of 25.4 days.
To account for detector asymmetries and seasonal background variations affecting
searches at timescales of tens of days or less we used a method of background estimation
described in~\cite{bazo2009icrc}.
The strongly reduced background in the search window led to a re-optimized bin radius of $4^{\circ}$.
No events were found during the selected periods. We derived a limit on the neutrino fluence of $3.7\!\times\!10^{-3}$~erg~cm$^{-2}$ in the energy range $8.3\!\times\!10^5$ to $7.6\!\times\!10^7$ GeV, based on the model spectrum from~\cite{reimer2009hd} shown in Fig.~\ref{fig_lim}.
\\
The searches for local excesses of events depend on declination dependent background expectations from scrambled data. Statistical errors were taken into account for the calculation of limits with the method from~\cite{Conrad2002}.
Systematic uncertainties were estimated with respect to the effects on the sensitivity to an $E^{-2}$ muon neutrino spectrum. Variation of simulation parameters 
led to a change of $-8$/+17\% for upper limit values. This is mainly due to remaining uncertainties about the light propagation in the ice as discussed in \cite{ic22ps09,GRB0803}.
To estimate the influence of reconstruction biases for simulated events, we compared data from below the horizon with a simulation of atmospheric neutrinos and derived an effect of $-9$/+4\% on the sensitivity.
Together with theoretical uncertainties for interaction and propagation of muons and neutrinos ($\pm5$\%), the total uncertainty adds up to $-13$/+18\% and was included in all upper limits and sensitivities with the method from~\cite{Conrad2002}.
\\
We presented a new approach to extend neutrino point source searches above the horizon, which increases the reach of neutrino telescopes beyond PeV energies.
The method was applied to IceCube data taken with 22 strings.
Results are consistent with the background-only hypothesis.
The derived flux limits are the first to cover neutrino point sources in the southern hemisphere at and above PeV energies.\\ 
\newline
We acknowledge the support from the following agencies:
U.S. National Science Foundation-Office of Polar Program,
U.S. National Science Foundation-Physics Division,
University of Wisconsin Alumni Research Foundation,
U.S. Department of Energy, and National Energy Research Scientific Computing Center,
the Louisiana Optical Network Initiative (LONI) grid computing resources;
Swedish Research Council,
Swedish Polar Research Secretariat,
and Knut and Alice Wallenberg Foundation, Sweden;
German Ministry for Education and Research (BMBF),
Deutsche Forschungsgemeinschaft (DFG),
Research Department of Plasmas with Complex Interactions (Bochum), Germany;
Fund for Scientific Research (FNRS-FWO),
FWO Odysseus programme,
Flanders Institute to encourage scientific and technological research in industry (IWT),
Belgian Federal Science Policy Office (Belspo);
Marsden Fund, New Zealand;
M.~Ribordy acknowledges the support of the SNF (Switzerland);
A.~Kappes and A.~Gro{\ss} acknowledge support by the EU Marie Curie OIF Program;
J.~P.~Rodrigues acknowledge support by the Capes Foundation, Ministry of Education of Brazil.
EB thanks A.M. Atoyan, C. Dermer, A. Reimer for useful discussion and A. Marscher for providing multi-wavelength data on 3C279.

\bibliography{bibliography}

\end{document}